\documentclass[prl,preprint]{revtex4}
\usepackage{graphicx}

\def\>{\rangle}
\def\<{\langle}
\newcommand{\be}{\begin{equation}}
\newcommand{\ee}{\end{equation}}
\newcommand{\ba}{\begin{eqnarray}}
\newcommand{\ea}{\end{eqnarray}}

\usepackage{bm}
\usepackage[dvips]{color}

\begin{document}

\title{Modeling  an adiabatic quantum computer}

\author{A. M. Zagoskin$^{1,2}$}
\author{S. Savel'ev$^{1,3}$}
\author{Franco Nori$^{1,4}$}
\affiliation{(1) Frontier Research
System, The Institute of Physical and Chemical Research (RIKEN),
Wako-shi, Saitama, Japan}
\affiliation{(2) Department of Physics and Astronomy, The University of British Columbia,  Vancouver, B.C., %V6T 1Z1 6224 Agricultural Rd.,
Canada}
\affiliation{(3) Department of Physics, Loughborough University, Loughborough, UK}
\affiliation{(4) Center for Theoretical Physics, CSCS, Department of
Physics, The University of Michigan, Ann Arbor, Michigan, %48109-1040
USA}

\begin{abstract}
We map   adiabatic quantum evolution   on the classical Hamiltonian dynamics of a 1D gas   (Pechukas gas)  and simulate the latter numerically.  This approach turns out to be both insightful and numerically efficient, as seen from our example of a CNOT gate simulation.   For a general class of Hamiltonians  we show that the   escape probability from  the initial  state scales no faster than   $|\dot{\lambda}|^{\gamma}$, where $|\dot{\lambda}|$ is the adiabaticity parameter. The scaling exponent for the escape probability  is $\gamma=1/2$ for all  levels, except the edge (bottom and top) ones, where $\gamma\lesssim 1/3$.  In principle, our method can solve arbitrarily large adiabatic quantum Hamiltonians.  
\end{abstract}

\maketitle

 Standard approaches  to quantum computing are based on applying a sequence of unitary operations to a multiqubit system, and the solution is encoded in an entangled superposition of its eigenstates, which is  fragile with respect to decoherence. Regrettably, there is a growing realization that this approach  is  not feasible in the near future. An alternative promising paradigm is adiabatic quantum computing \cite{Farhi2000,Farhi2001,Steffen2003} (AQC), where the solution is encoded in the ground state of the system  evolving   under an adiabatically slow change of a control parameter $\lambda$. Remaining in the ground state  protects the system against relaxation and dephasing \cite{Childs2002}.  

  In principle, any standard quantum circuit can be realized by an AQC \cite{Aharonov2004,Mizel2006}. Following the AQC approach, we reach the ground state of a complex Hamiltonian $H_0$, which encodes the solution to a given quantum algorithm, by adding to $H_0$ a large bias term, $ZH_b, \:\: Z\gg 1,$
\begin{equation}
H(\lambda(t)) = H_0 + \lambda(t) ZH_b, \label{eq_H0},
\end{equation}
such that $H(\lambda=1)$ has a non-degenerate, easily achievable ground state. 
The intrinsic limitation  imposed on AQC is the finite probability of excitation, via   Landau-Zener tunneling \cite{Landau2003,Ashhab2006},
 at any finite evolution speed (the adiabaticity parameter $|\dot{\lambda}|$). This tunneling remains even after the effects of external and thermal noise are eliminated.
Research on AQC has so far concentrated on evaluating and minimizing the probability   of leaving the ground state. Both polynomial \cite{Hogg2003,Boulatov2005,Mizel2006} and exponential \cite{Mitchell2005,Znidaric2005} slowdown was predicted using AQC, and this important  question remains open.

In this paper we investigate what is the probability of escape from the ground state, and how far, {\em on average}, a system can deviate from the ground state   during the  adiabatic evolution.  We do this by mapping the parametric evolution of the system (\ref{eq_H0}) on the classical Hamiltonian dynamics of a 1D gas model with long-range repulsion (Pechukas gas \cite{Pechukas1983,Haake2001}) and simulating the latter numerically.  
This approach, which had not been used so far in the field of AQC, turns out to be both physically insightful and numerically efficient way to solve (\ref{eq_H0}). For a general class of Hamiltonians, the probability to stay in the same state shows a universal power-law dependence for all the energy levels except for the ground and top excited states, and this difference can be qualitatively understood within the Pechukas gas model.
We also develop a kinetic theory that could, in principle, solve arbitrarily large adiabatic Hamiltonians.

{\em  Pechukas gas dynamics.---\/} 
In this approach we consider the instantaneous eigenstates, $|n\rangle(\lambda)$, and eigenvalues, $E_n(\lambda)$, of the Hamiltonian (\ref{eq_H0}): $H(\lambda)|n\rangle  = E_n(\lambda)|n\rangle$. The Hamiltonian $H(\lambda)$ is fully determined by the set of all its instantaneous matrix elements, $\langle m|H(\lambda)|n\rangle.$ The latter can be determined from the closed set of differential equations (see Refs.~\cite{Pechukas1983,Haake2001}):
\begin{eqnarray}\label{eq_Pechukas1}
\frac{d}{d\lambda}{x}_m = v_m   ;\:\:
\frac{d}{d\lambda}v_m =  2\sum_{m\ne
n}\frac{|l_{mn}|^2}{(x_m-x_n)^3}  ;\:\:
\frac{d}{d\lambda}{l}_{mn} =   \sum_{k\neq m,n}   l_{mk}\: l_{kn} \left(\frac{1}{(x_m-x_k)^2}-\frac{1}{(x_k-x_n)^2}\right), 
\end{eqnarray}
where $x_n(\lambda) = E_n(\lambda)$,  $v_n(\lambda) = \langle n| ZH_b|n\rangle$, and $l_{mn}(\lambda) = (E_m(\lambda)-E_n(\lambda)) \langle m| ZH_b|n\rangle$.
Equations (\ref{eq_Pechukas1}) describe the classical Hamiltonian dynamics of a 1D gas with repulsion, where $\lambda$ plays the role of time, and the $n$th ``particle" has a position $x_n(\lambda)$ and velocity $v_n(\lambda)$. The particle-particle repulsion is determined by the   ``relative angular momenta" $l_{mn}(\lambda)$. 

The mapping of AQC to the classical dynamics of Eq.~(\ref{eq_Pechukas1}) is exact and applies to any Hamiltonian.
All the information about the Hamiltonian $H_0$ is contained in the initial values, $x_m(\lambda=1),\: v_m(\lambda=1),$ and $l_{mn}(\lambda=1)$. Those are given by the appropriate matrix elements of the Hamiltonian (\ref{eq_H0}) at $\lambda=1$, $H(1) = H_0 +  ZH_b \equiv Z(H_b + Z^{-1}H_0)$. By choice, $H(1)$ has non-degenerate, well separated levels, with an easily reachable ground state. The initial conditions for Eq.~(\ref{eq_Pechukas1}) can be obtained perturbatively in $Z^{-1}$ to any accuracy.

{\em CNOT gate.---\/} Consider a specific example of an AQC. An arbitrary $N$-step, $M$-qubit quantum algorithm can be encoded in $H_0$ using the ``ground state quantum computing" approach \cite{Mizel2001}, whereby every qubit is represented by an array of $2\times(N+1)$ quantum dots sharing a single electron. The state of the $m$th qubit on  the  $(n+1)$st step of the algorithm is given  by the probability amplitude to find the electron on either the quantum dot $(m,n,0)$ or $(m,n,1)$. The solution is determined by the ground state probability amplitudes on the quantum dots $(m,N,0)$  and $(m,N,1)$.

For a simple universal quantum gate, the CNOT (where $N=1,\:M=2$), $H_0$ can be written as \cite{Mizel2001,Mizel2006} 
\begin{eqnarray}\label{eq_CNOT}
H_{\rm CNOT} = (c^{\dag}_{010}C^{\dag}_{11} - c^{\dag}_{000}C^{\dag}_{10})(C_{11}c_{010} - C_{10}c_{000})+ \nonumber\\
(c^{\dag}_{011}C^{\dag}_{11} - c^{\dag}_{001}C^{\dag}_{10}\sigma^x)(C_{11}c_{011} - \sigma^xC_{10}c_{001})+ 
C^{\dag}_{00}C_{00}C^{\dag}_{11}C_{11}+C^{\dag}_{01}C_{01}C^{\dag}_{10}C_{10},
\end{eqnarray}
where $c^{\dag}_{mnj}$ creates an electron on the  corresponding dot, and 
$C^{\dag}_{mn} = (c^{\dag}_{mn0},  c^{\dag}_{mn1})$.  The ground state energy of (\ref{eq_CNOT}) is zero. 
In order to specify the initial state of the qubits before the operation, a small correction must be added.

\begin{figure}[b]
\includegraphics[width=8.8cm]{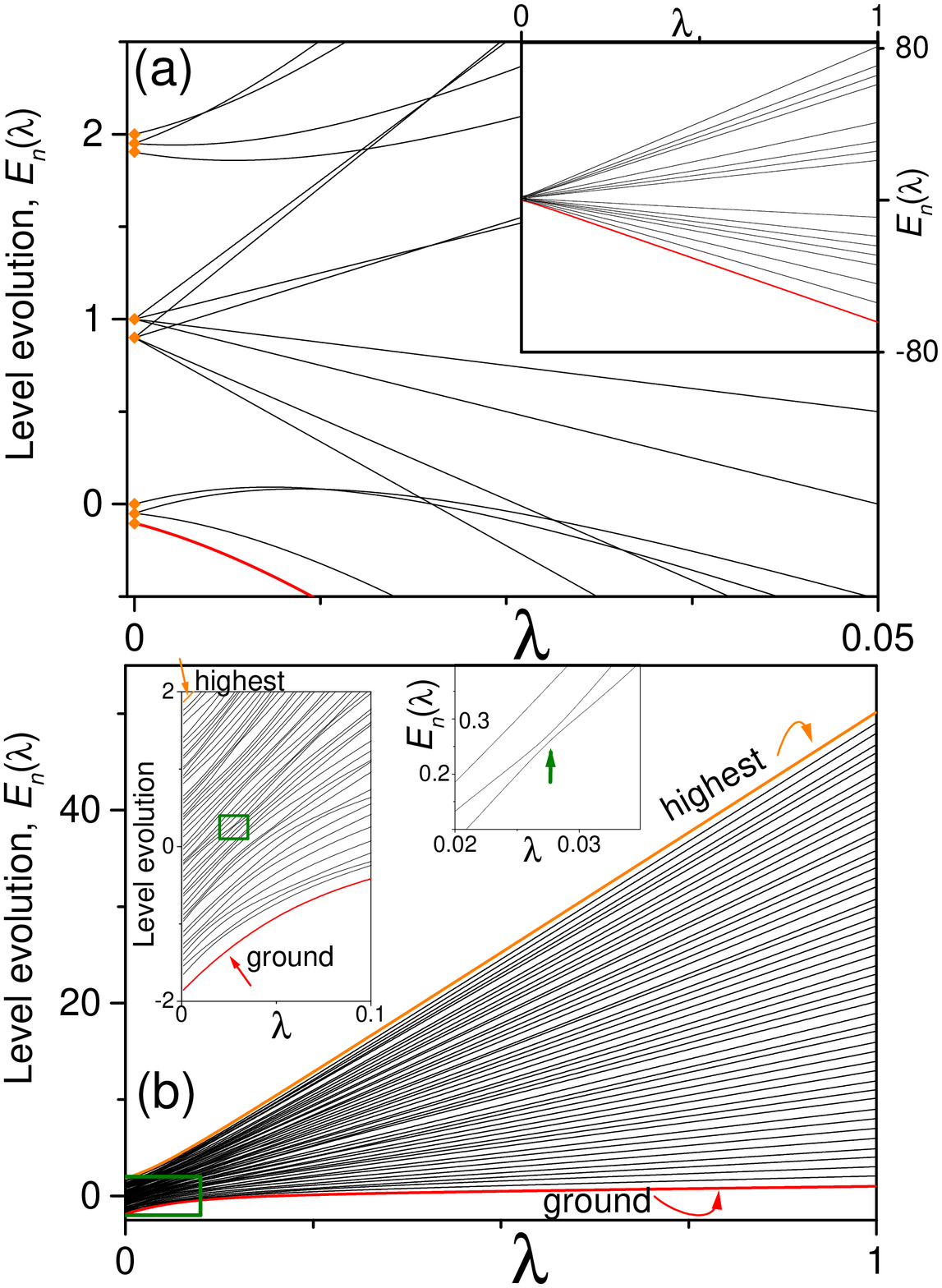}
\caption{(Color online) (a) CNOT gate simulation for the operation $|00\rangle \to |00\rangle$, described by the Hamiltonian (\ref{eq_H0}) with $H_0 = H_{\rm CNOT} + \delta H$ (see Eq.~(\ref{eq_CNOT}) and below). Some levels cross due to symmetry, but not the ground state. The small orange diamonds on the vertical axis of the main panel  show the results of direct diagonalization of $H_0$ and coincide with the results of the evolution (\ref{eq_Pechukas1}) to   four  significant figures. 
(b) Evolution of 50 energy levels in Eq.~(\ref{eq_Pechukas1}) as a function of the parameter $\lambda$,  for a single GUE realization of the  Hamiltonian (see text). The energy scale is determined by imposing the normalization $\langle \Delta E \rangle = \hbar|\dot{\lambda}|,$ where $\langle \Delta E \rangle = \langle  E_{n+1} - E_n\rangle$ is the average level spacing.  The r.m.s. amplitude of the fluctuations in $E_n$ and $\langle m| ZH_b|n\rangle$ is $0.1 \langle \Delta E \rangle $. 
  The ground state (red) and the highest excited state (orange) are  labeled. It is clearly seen in the insets (which enlarge the green boxes) that avoided level crossings appear at small values of $\lambda$ (e.g., green vertical arrow in the top inset). In the left inset notice the significant separation between the ground state and the first excited state.
}\label{fig1}
\end{figure}

The results shown in Fig.~\ref{fig1}a correspond to CNOT   $|00\rangle\to |00\rangle.$ To achieve this, the term $\delta H = \varepsilon (c^{\dag}_{000}c_{000} + c^{\dag}_{100}c_{100})$, with $\varepsilon=-0.1$, was added to $H_0$. Remarkably, even though the initial conditions for the Pechukas equations (\ref{eq_Pechukas1}) were only calculated to first order in $Z^{-1} = 0.1$, the results agree with the exact diagonalization to four significant figures, indicating a high efficiency of the approach. The same accuracy holds for the CNOT acting on all the other basis states. All the degeneracies of the Hamiltonian (\ref{eq_CNOT}) are precisely reproduced in our approach.  Since any 1D gas with repulsive interactions naturally expands, the decrease of the bias potential $\lambda ZH_b$ corresponds to a  contraction from $H$ to $H_0$.  Such evolution   can be considered a  Loschmidtian time reversal \cite{Klimontovich1986} of the natural expansion of the Pechukas gas.
Note that while the levels generally repel, certain groups of levels can cross, due to the symmetries of the Hamiltonian (\ref{eq_CNOT}). 

{\em Statistics of level occupation.---\/} Unlike the simple example of a CNOT gate presented above, when considering a general AQC case, we would benefit from the knowledge of the statistical behaviour, for a given class of AQC problems (i.e., Hamiltonians $H_0$).   An ensemble of Hamiltonians $H_0$ corresponds to a distribution of  initial conditions in  the Pechukas dynamics in Eq.~(\ref{eq_Pechukas1}), over which an appropriate averaging must be taken. We choose a set  described by one of the random matrix theory (RMT) Gaussian ensembles \cite{Guhr1998}. 
  This, in particular, means that the  distribution of $x_{n}(\lambda=1), v_{n}(\lambda=1)$ and $l_{mn}(\lambda=1)$ is Gaussian. Such an assumption about the behaviour of a large collection of qubits with varied couplings is   reasonable and was used recently in \cite{Boulatov2005,Mitchell2005}, but must  still be taken {\em cum grano salis}. For example, the numerical calculations in Ref.~\onlinecite{Znidaric2005} show that for the 3-SAT problem, RMT describes well only the bulk of the spectrum.   
In the case of, e.g., flux qubits threaded by a magnetic field, the natural choice is the Gaussian unitary ensemble (GUE) of general Hermitian matrices. (If the system has time-reversal symmetry, one should  instead use the Gaussian orthogonal ensemble (GOE) of real Hamiltonians \cite{Guhr1998}.) Note that  the solutions of (\ref{eq_Pechukas1}) are described by RMT only when $\lambda \to 0$. For  most of the evolution, the perturbation $ZH_b$ will impose its own symmetry.  
  Eventually, we are interested in the probabilities,
$P(m|n),$ for the system to end in the  eigenstate $|m\rangle$ at $\lambda=0$ after starting in the eigenstate $|n\rangle$ at $\lambda=1$. 

In the absence of external and thermal noise, the interlevel transitions are due to the   Landau-Zener tunneling.
 At each level anticrossing, where two levels approach to a minimum separation $\Delta_{\rm min},$ they may exchange their current occupations: $P_{\lambda}(m|n) \longleftrightarrow P_{\lambda}(m+1|n)$. The probability of this process   \cite{Landau2003}
\begin{equation}
p_{LZ} = \exp\left(-\frac{\Delta_{\rm min}^2}{4\pi\hbar|\langle m| ZH_b|m+1\rangle||\dot{\lambda}|}\right) \label{eq_LZ}
\end{equation}
 strongly depends on $|\dot{\lambda}|$. We always assume a uniform evolution, $|\dot{\lambda}| = 1/T$.

\begin{figure}[floatfix]
\includegraphics[width=8.8cm]{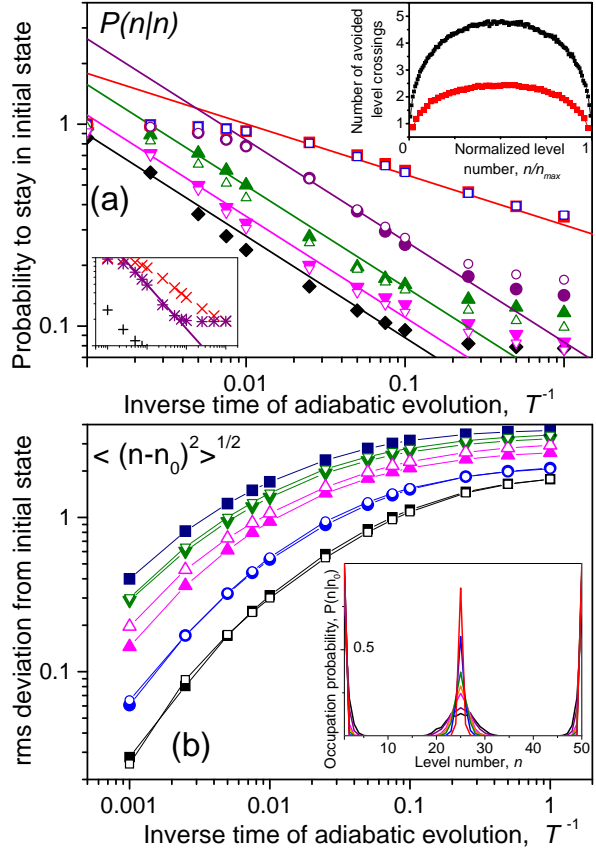}
\caption{(Color online) (a) Probability $P(n|n)$ for the system to remain in the initial state $|n\rangle$, as a function of $1/T$,   during the adiabatic evolution (i.e., decreasing the external field with   constant speed $\left|\dot{\lambda}\right|=1/T$). The data are averaged over 400  GUE realizations with $N=50$. The power-law dependence $P \propto  T^{\gamma}$ is clearly seen, with two distinct exponents: (i) $\gamma=1/2$ for all the bulk states and (ii) $\gamma \lesssim 1/3$ for the edge states. For example, for $N=50$: $\gamma=1/2$ for $n=2\:(49)$, solid (empty) purple circles;   $n=5\:(45)$, solid (empty) green triangles;   $n=10\:(40)$, solid (empty) upturned magenta triangles; $n=25$, black diamonds; but  $\gamma=1/4$ for the ground state ($n=1$, red squares) and the highest excited state ($n=50$, empty blue squares). For $N=150$ (left inset):   $\gamma=1/2$ for $n=2$ (purple stars), $n=75$ (black pluses); but $\gamma\approx 1/3$ for the ground state ($n=1$, red crosses).  However, the average number of avoided level crossings (right inset) is a smooth function of the level number for both $N=50$ (red) and $N=150$ (black). 
 As expected in the absence of  external noise, the probability $P(n|n)$ saturates at 1 as $1/T \to 0$ (when $1/T < 0.01$). 
 (b) The standard deviation $\langle(n-n_0)^2\rangle$ of the system from the initial state $n=n_0$ during the adiabatic evolution, as a function of $1/T$ $(N=50)$. The scaling is approximately power-law, but with the exponent smoothly dependent on the initial state. (Inset) Probability $P(n|n_0)$ to occupy level $|n\rangle$ at the end of the evolution, starting from level $n_0 = 1, 25, 50$ ($N=50$). Different curves correspond (top to bottom peaks) to $10^3 \times |\dot{\lambda}|$ =  1,  2.5, 5, 10, 25, 50, 75. }\label{fig2}
\end{figure}

 The results of simulations for the GUE are shown in Figs.~\ref{fig1},\ref{fig2}. Remarkably,  the probability to remain in the initial state by the end of the process  scales  as $T^{1/2}$, with the {\em same power}, 1/2,   for   {\em all} the levels, {\em except} the edge ones (top and bottom eigenenergies). For the edge levels the probability scales as $T^{1/4}$ in the case of 50 levels, and as $\approx T^{1/3}$ for the case of 150 levels. 
 
 The different scaling exponent of the edge levels, compared to the bulk,  cannot be only due to the long range cubic repulsion in Eq.~(\ref{eq_Pechukas1}) between the levels. The latter is obviously responsible for the difference between the bulk and the margin of the spectrum in the number of anticrossings (i.e., collisions in the 1D gas), but this dependence is a smooth function of the level number.
It is the  edge (top or bottom) position of the level, that is important.
Indeed, while the separation of the internal levels after a collision is limited by the subsequent collisions with the levels on the opposite side, this does not apply to the ground state and the top state. This can be clearly seen in the top right inset of   Fig.~\ref{fig2}a. One also sees there that the number of collisions (i.e., avoided crossings) for these edge levels is on average less than one. This stresses  the special robustness of the ground  and top states, precisely due to their edge positions. 
 
 Another remarkable result is  the behaviour of the r.m.s. deviation of the system from its initial state (Fig.~\ref{fig2}b). It also scales no faster than a power of $T$. However,  the exponent here is a {\em smooth} function of the level number $n$. This means that the observed behaviour is not described by a simple diffusion equation.

{\em Kinetic theory for large adiabatic quantum Hamiltonians.---\/} The above results indicate that the approximate ground state energy of a system described by the GUE can be efficiently determined by adiabatic quantum evolution. Indeed, to reach the accuracy $\epsilon$, the adiabaticity parameter must scale as a power of $\epsilon$. On the other hand, e.g.,    the time necessary to find the ground state energy of a complex system (e.g., a spin glass) with classical annealing algorithms generally depends on $1/\epsilon$ exponentially (see Ref. \cite{Santoro2002} and references therein). 
It is therefore tempting to conclude that the adiabatic evolution of a quantum computer could provide an exponential speedup, on average, for this problem.

 In order to confirm this conjecture and establish the limits of its validity, we need to consider larger systems. Then the brute force approach to solving the set (\ref{eq_Pechukas1}) becomes inefficient.
   We can instead identically rewrite it  as a chain of equations (see e.g.~\cite{Klimontovich1986}) for the microscopic distribution functions $F_1(x,v,n) = \sum_{j} \delta(x-x_j)\delta(v-v_j)\delta(n-n_j),$
$F_2(x,v,n;y,u,m;l) = \sum_{j,k} \delta(x-x_j)\delta(v-v_j)\delta(n-n_j)\delta(y-x_k)\delta(u-v_k)\delta(m-n_k)\delta(l-l_{jk}),$
$G_1(l) = \sum_{jk} \delta(l-l_{jk}),$ $\dots$  Here we included the occupation numbers of the levels in the description. After ensemble averaging, these produce the BBGKY (Bogoliubov-Born-Green-Kirkwood-Yvon) chain \cite{Klimontovich1986} for the averaged distribution functions, $f_1 = \langle F_1 \rangle, g_1 = \langle G_1 \rangle\dots$. 
The first equation of this chain is 
\begin{equation} 
\left[\frac{\partial }{\partial \lambda} + v\frac{\partial}{\partial x}\right]f_1(x,v,n) = 
2 \frac{\partial}{\partial v} \sum_m \int dl\:dy\:du \frac{|l|^2}{(y-x)^3}f_2(x,v,n;y,u,m;l). \label{eq_kinetic}
\end{equation}

Following the standard kinetic approach \cite{Klimontovich1986}, the   BBGKY hierarchy can be truncated by representing the r.h.s.~of (\ref{eq_kinetic}) as a  term responsible for the long-range interactions (where $f_2$ can be factorized, $f_2(x,v,n;y,u,m;l) \to f_1(x,v,n) f_1(y,u,m) g_1(l)$), plus the collision integral, $I_{\rm St}$. This leads to an approximate equation for $f_1$: 
\begin{equation}
\left[\frac{\partial }{\partial \lambda} + v\frac{\partial}{\partial x} - 2 \Gamma \left(\sum_m {\cal P}\!\!\int \:dy\:du\: \frac{f_1(y,u,m)}{(y-x)^3}\right)\frac{\partial}{\partial v}\right]f_1(x,v,n) = 
I_{\rm St},  \label{eq_kinetic2}
\end{equation}
where the effective repulsion $\Gamma = \int dl\; g_1(l)\; |l|^2$, ${\cal P}$ denotes the principal value of an integral,  and the collision integral 
\begin{equation} 
I_{\rm St} = 2 \Gamma_{\rm St}\:\sum_m \int du \: (u-v) \: p_{LZ}(u-v)\: \left[ f_1(x,v,m)f_1(x,u,n)-f_1(x,v,n)f_1(x,u,m)\right].  \label{eq_stoss}
\end{equation}
Here $\Gamma_{\rm St}$ is a constant; the collision integral (\ref{eq_stoss}) describes the population exchange of two anticrossing levels due to Landau-Zener tunneling (\ref{eq_LZ}).

If we are only interested in the behaviour of energy levels, but not in their occupation (i.e. considering $\tilde{f}_1 = \sum_n f_1$), the kinetic equation for $\tilde{f}_1$ will have the form (\ref{eq_kinetic2}), but without the collision integral.
When the full hierarchy of dynamical equations (\ref{eq_Pechukas1}) becomes intractable due to the large size of the system, one can truncate it to only a few lower states. The influence of the higher-energy states can be then taken into account statistically using the kinetic equation for $\tilde{f}_1$.
 Due to the long-range interactions in the system, one expects that the average behaviour of the low-lying states can be thus accurately predicted. Thus this approach could be extended to simulate any large quantum system.

{\em Conclusions.---\/} We   model  the   quantum adiabatic evolution of any Hamiltonian by mapping it to the dynamics of a 1D gas. Using a CNOT gate as an example, we found that the approach is reliable.   In the case of GUE Hamiltonians, we found that the behaviour of the system is not described by a simple diffusion equation, because the standard deviation from the initial state and the probability to stay in it scale differently with the adiabaticity parameter $|\dot{\lambda}|$. The probability to stay in the same state shows a universal power-law dependence for all the levels except for the ground and top excited states, and this difference can be qualitatively understood within the Pechukas gas model. For the investigation of large systems, where direct simulation is impractical, we propose a kinetic approach based on the BBGKY chain for the Pechukas gas. The results of this paper indicate that an adiabatic quantum evolution can provide an  exponential, on average, speedup compared to the classical simulated annealing in finding an approximate ground state energy of a complex system.

  We acknowledge partial support from the NSA, LPS, ARO, NSF grant
No. EIA-0130383, JSPS-RFBR 06-02-91200, MEXT
Grant-in-Aid No. 18740224, EPSRC via No. EP/D072581/1 and the NSERC Discovery Grants Program (Canada).

\end{document}